\documentclass[authoryear,preprint,3p,12pt]{elsarticle}

\usepackage{amssymb}
\usepackage{amsmath}
\usepackage{graphicx}
\usepackage{caption}
\usepackage{subcaption}
\usepackage{float}
\usepackage{color}
\usepackage{multicol}

\journal{Optics Communication}

\begin{document}

\begin{frontmatter}



\title{Coupled Mode Theory of Microtoroidal Resonators with a One-dimensional Waveguide}

\author[CEA]{Thi Phuc Tan Nguyen}
\ead{phuctan3108@gmail.com}
\author[ASTAR]{Leonid Krivitsky}
\author[CQT,NUSPhys,NTUIAS,NIE,Maju]{Leong Chuan Kwek}

\address[CEA]{SyMMES/INAC/CEA, 17 Rue des Martyrs, F-38054 Grenoble Cedex 9, Grenoble, France}
\address[CQT]{Centre for Quantum Technologies, National University of Singapore, 3 Science Drive 2, Singapore 117543}
\address[NUSPhys]{Physics Department, National University of Singapore, 2 Science Drive 3, Singapore 117551}
\address[ASTAR]{Data Storage Institute, Agency for Science, Technology and Research (A*STAR), 2 Fusionopolis Way, $\#$08-01 Innovis, Singapore 138634}
\address[NTUIAS]{Institute of Advanced Studies, Nanyang Technological University,
60 Nanyang View, Singapore 639673, Singapore}
\address[NIE]{National Institute of Education, Nanyang Technological University,
1 Nanyang Walk, Singapore 637616, Singapore}
\address[Maju]{MajuLab, CNRS-UNS-NUS-NTU International Joint Research Unit, UMI 3654, Singapore}


\begin{abstract}
	We study the transmission of light through a system consisting of an arbitrary number $N$ of microtoroidal resonators coupled to a one-dimensional (1D) waveguide. The transmission $T$ through such a system and its full-width at half-maximum (FWHM) are calculated for various values of $N$ and mutual-mode coupling coefficients. We found that at small mutual-mode coupling, the minimum transmission vanishes exponentially with $N$ while the FWHM is proportional to $\sqrt{N}$. At big mutual-mode coupling, as the number of resonators increases, the mode-splitting is reduced. Our findings contribute to better understanding of novel interfaces between quantum emitters and resonant photonic structures for quantum information processing.
\end{abstract}

\begin{keyword}


	Microtoroidal resonator, coupled-mode theory, transmission spectrum
\end{keyword}

\end{frontmatter}



	\section{Introduction}


	The atom-photon coupling can be greatly enhanced by using a cavity \cite{AtomPhotonTannoudji}. The strong coupling between atoms and photons is one of the main building blocks for quantum-state transfer \cite{10.1038/nature11023} and leads to a number of interesting quantum effects \cite{ExploreQuantum}. Some examples of these effects are the vacuum-Rabi oscillation \cite{PhysRevLett.94.033002, PhysRevLett.101.223601} and tunable photon transmission in 1D waveguides \cite{PhysRevA.90.053822}.

	Nevertheless, attaining the strong coupling regime remains technically challenging due in part to the need for high quality-factor (Q-factor) cavities \cite{RevModPhys.87.1379}. One of the several promising approaches for fabricating high Q-factor cavities is based on the use of integrated silicon photonics platforms. These platforms combine the benefits of intrinsically stable operation, CMOS compatible fabrication and compact footprint. A considerable amount of efforts has been made to achieve the highest possible enhancement of atom-photon interaction by optimizing various designs for silicon-based cavities. Photonic crystal cavities (PCCs) reaching experimental values of Q-factors at the order $10^3$ have been reported \cite{KassaBaghdouche20153467, PhysRevLett.102.173902}. In some cases, a Q-factor as high as $10^5$ has also been achieved \cite{10.1038/srep26038}. Other cavity designs such as microtoroidal cavities can reach very high values of Q-factor ($>10^5$) \cite{SiNMicroringGoykhmanBorisUriel, Hosseini:09}. They are potentially useful for achieving strong coupling between quantum emitters and cavity modes \cite{PhysRevLett.102.083601}.

	The quantum-mechanical description of the coupling between an emitter and a photonics platform can be studied using an input-output theory \cite{PhysRevA.31.3761}. If there is no quantum emitter and the input light is coherent, a classical calculations can be employed. These calculations are typically done with the coupled-mode theory (CMT) \cite{CMT@HHaus}. Using the CMT, the transmission spectrum for one or two microtoroidal resonators coupled to four input-output ports has been reported \cite{784592}. 
In the presence of fabrication and dielectric defects for instance, the two counter-propagating modes inside a ring resonator can couple to each other \cite{Little:97,Barwicz:04}.
When mutual-mode coupling is strong, it has been shown that the mode splitting occurs in the transmission spectrum $T$ \cite{Zhang:08}. This mode splitting reduces the coupling efficiency between the cavity modes and the quantum emitters at the resonant frequency of the cavity.
	
	In this paper, we present a theoretical study of the transmission of light in a 1D waveguide via a system of $N$ microtoroidal resonators with mutual-mode coupling. There are two main motivations behind our research. On the one hand, we would like to understand the effects that an array of $N$ microtoroidal resonators has on the mode-splitting. We discovered that the problem of mode-splitting is reduced for large $N$. We discuss in section 3 the dependence of the depth and the width of the transmission spectrum T on $N$. On the other hand, the research on various systems of microtoroidal resonators has been mainly experimental or numerical \cite{nphoton.2009.237, PhysRevLett.99.173603, 2040-8986-12-10-104008}. Theoretical work has been mostly focused on one or two resonators \cite{OE.18.008367, 784592, Zhang:08} or with only one mode present in each ring \cite{CaltechTHESIS:05242012-174045229}. A theory for the case of an arbitrary number of ring resonators that  have mutual-mode coupling and that are coupled to a one-dimensional (1D) waveguide has not been investigated to the best of our knowledge. A more systematic framework, which simplifies the study for any number of resonators, is useful for future progress of the field.

	The paper is organised as follows. In section \ref{SecMethods}, we describe the general framework that allow us to study  the problem of arbitrary $N$ microtoroidal resonators and derive the $N$-ring transfer matrix. 
After giving the formal expression for the transmission spectrum in section \ref{SecRes&Dis}, we show the numerical results for $T$ and discuss its dependence on $N$. We summarize our results in section \ref{SecConclusion} with some remarks regarding the future directions.


	\section{Methods}
	\label{SecMethods}

	\begin{figure}[h]
		\centering
		\includegraphics[scale=0.5]{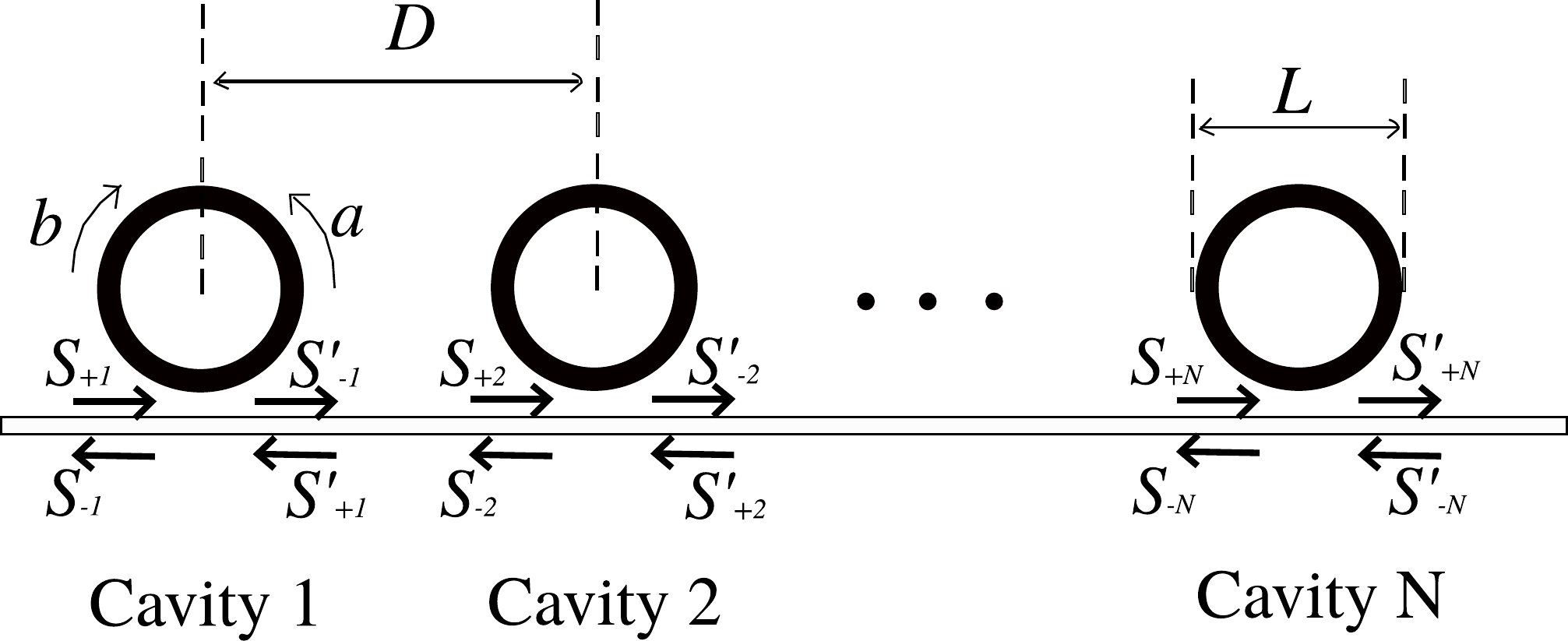}
		\caption{Arrangement of N microtoroidal resonators (black circles) along a waveguide (transparent horizontal line); $a$, $b$, the two modes of a resonator; $D$, distance between two adjacent resonators; $L$, radius of a resonator; $S_{+/- i} (S'_{+/- i})$, incoming/outgoing light field from the left (right) \label{NRing_depiction}}
	\end{figure}
	We begin our analysis by considering $N$ identical microtoroidal resonators with radii $L$  coupled to a 1D waveguide as depicted in Fig. \ref{NRing_depiction}. Let $D$ be the distance between the $i^{\text{th}}$ and $(i+1)^{\text{th}}$ rings. For simplicity, we assume that all the resonators have the same radii and that they are placed along the waveguide such that the distances between any two adjacent rings are equal. The case in which $D$ varies among the resonators is discussed in Section \ref{SecRes&Dis}.
	
	Let us consider the $i^{\text{th}}$ resonator. Let $S_{+/- i}$ and $S'_{+/- i}$ denote the incoming/outgoing light field from the left and the right respectively. The two counter-propagating modes that are denoted $a_i$ and $b_i$ oscillate at frequencies $\omega_{a_i}$ and $\omega_{b_i}$, respectively. $a_i$ and $b_i$ are coupled to the waveguide with coupling coefficients $\kappa_{a_i}$ and $\kappa_{b_i}$ given below
	\begin{equation}
		\kappa_{a_i} = \sqrt{ \frac{\omega_{a_i}}{Q_{i,e}} }, \ \kappa_{b_i} = \sqrt{ \frac{\omega_{b_i}}{Q_{i,e}} },
	\end{equation}
where $Q_{i,e}$ are the quality factors of the resonators. The coupling of the resonators to the waveguide gives rise to the decay rates $\Gamma_{a_i,e} = \frac{\omega_{a_i}}{2 Q_{i,e}}$ and $\Gamma_{b_i,e} = \frac{\omega_{b_i}}{2 Q_{i,e}}$ of the cavity modes. The coupling to other lossy channels leads to the intrinsic decay of $a_i$ and $b_i$ at the rate $\Gamma_{a_i,i}$ and $\Gamma_{b_i,i}$, respectively. Thus, the total decay rates are
	\begin{equation}
		\Gamma_{a_i} = \Gamma_{a_i,e} + \Gamma_{a_i,i}, \ \Gamma_{b_i} = \Gamma_{b_i,e} + \Gamma_{b_i,i}
	\end{equation}
We also take into account the coupling between $a_i$ and $b_i$. This coupling is characterised by the coefficients $u_i$ that are usually taken to be real \cite{Zhang:08,4675187} since the coupling is essentially the energy transfer between $a_i$ and $b_i$ without losses. One of the most common reasons for $u_i$ to be non-zero is dielectric defects. We further let $u_i$ to be frequency-independent as we work in linear optics regime, though the current framework can be extended to $\omega$-dependent coupling.


		\subsection{Coupled-mode analysis}
		\label{SubSecCMA}
In this subsection, we solve for $a_i$ and $b_i$ using the CMT \cite{CMT@HHaus} in the frequency domain. On the  one hand, this approach gives the same results as the steady-state solution from the time-domain analysis for the case in which $N=1$ and the system is pumped from the left with $S_{0} e^{i \omega t}$. On the other hand, $a_i(\omega)$ and $b_i(\omega)$ can be obtained by solving algebraic equations while $a_i(t)$ and $b_i(t)$ are the Fourier transforms of the former.
		
By making Fourier transform the coupled-mode equations for the $i^{\text{th}}$ resonator \cite{Zhang:08}, the set of equations for $a_i$ and $b_i$ in the $\omega$-domain is
		\begin{equation}
			\left\{
			\begin{array}{c}
				\left[ i ( \omega - \omega_{a_i} ) + \Gamma_{a_i} \right] a_i(\omega) + i u_i b_i(\omega) = -i \kappa_{a_i} S_{+i}(\omega) \\
				\\
				\left[ i ( \omega - \omega_{b_i} ) + \Gamma_{b_i} \right] b_i(\omega) + i u_i a_i(\omega) = -i \kappa_{b_i} S'_{+i}(\omega)
			\end{array} 
			\right.
		\end{equation}
To simplify the subsequent analysis, we employ the following shorthand notations
		\begin{equation}
			A_i(\omega) = i ( \omega - \omega_{a_i} ) + \Gamma_{a_i}, \ B_i(\omega) = i ( \omega - \omega_{b_i} ) + \Gamma_{b_i},
		\end{equation}
		\begin{equation}
			D_i(\omega) = A_i(\omega) B_i(\omega) + u_i^2,
		\end{equation}
		and
		\begin{equation}
			t_{A_i} = 1 - \frac{ |\kappa_{b_i}|^2 A_i(\omega) }{ D(\omega) }, \ t_{B_i} = 1 - \frac{ |\kappa_{a_i}|^2 B_i(\omega) }{ D(\omega) } .
		\end{equation}
		
It is straightforward to show that the solutions $a_i(\omega)$ and $b_i(\omega)$ with respect to $S_{+i}(\omega)$ and $S'_{+i}(\omega)$ are
		\begin{align}
			a_i(\omega) & = \frac{ -i \kappa_{a_i} B_i(\omega) }{ D_i(\omega) } S_{+i}(\omega) - \frac{ u_i \kappa_{b_i}}{ D_i(\omega) } S'_{+i}(\omega) \label{a-S_in-Relation} \\
			b_i(\omega) & = \frac{ -i \kappa_{b_i} A_i(\omega) }{ D_i(\omega) } S'_{+i}(\omega) - \frac{ u_i \kappa_{a_i}}{ D_i(\omega) } S_{+i}(\omega) \label{b-S_in-Relation} 
		\end{align}


		\subsection{Input-output relation}
		\label{InOutRel}
In this subsection, we relate the signal on the left most $S_{+/- 1}$ to the right most $S'_{+/- N}$ side of a system of an arbitrary number $N$ of resonators. In other words, we want to find the matrix $T^{(N)}$ that satisfies the following condition:
\begin{equation}
			\left( \begin{array}{c} S'_{- N} \\ S'_{+ N} \end{array} \right) = T^{(N)} \left( \begin{array}{c} S_{+ 1} \\ S_{- 1} \end{array} \right).
		\end{equation}
		
We divide the problem of finding $T^{(N)}$ into two smaller ones: (i) obtaining the single-ring transfer matrix $T_{i}$ and (ii) determining the relation of the signal between two adjacent resonators.
\linebreak

(i) \textit{Transfer matrix $T_i$}:

The transfer matrix relates the left port of the $i^{\text{th}}$ resonator to its right port. It is defined as
		\begin{equation}
			\left( \begin{array}{c} S'_{-i} \\ S'_{+i} \end{array} \right) = T_i \left( \begin{array}{c} S_{+i} \\ S_{-i} \end{array} \right)
		\end{equation}
		
By using $a_i(\omega)$ and $b_i(\omega)$ in equations (\ref{a-S_in-Relation}) and (\ref{b-S_in-Relation}) together with the following relations between the cavity modes and the signal
		\begin{equation}
		\left\{
		\begin{array}{c}
			S'_{-i} = e^{-i \Phi_L} \left( S_{+i} - i \kappa_{a_i}^* a_i \right) \\
			S_{-i} = e^{-i \Phi_L} \left( S'_{+i} - i \kappa_{b_i}^* b_i \right) 
		\end{array}
		\right. ,
		\end{equation}
the transfer matrix for the $i^{\text{th}}$ ring is given by
		\begin{equation}
		\label{T_i}
			T_i = \left( \begin{array}{c} e^{-i \Phi_L} \left( t_{B_i}(\omega) + \frac{u_i^2 |\kappa_{a_i}|^2 |\kappa_{b_i}|^2}{D_i(\omega)^2 t_{A_i}(\omega)} \right) \\ \\ -i \frac{ u_i \kappa_{b_i}^* \kappa_{a_i} }{D_i(\omega) t_{A_i}(\omega)} \end{array}  \begin{array}{c} i \frac{ u_i \kappa_{a_i}^* \kappa_{b_i} }{D_i(\omega) t_{A_i}(\omega)} \\  \\ e^{i \Phi_L} \frac{1}{t_{A_i}(\omega)} \end{array} \right) .
		\end{equation}
		
	We note that in the above formula, $\Phi_L$ is just the phase factor that the light picks up when travelling across the $i^{\text{th}}$ microtoroidal resonator.

(ii) \textit{Phase relation for light between two resonators}:

When the light propagates along the waveguide from one ring to the next, it  accumulates a phase. Thus, we have the following phase relation
		\begin{equation}
			\left( \begin{array}{c} S_{+(i+1)} \\ S_{- (i+1)} \end{array} \right) = T^{ph}_{i+1, i} \left( \begin{array}{c} S'_{-i} \\ S'_{+i} \end{array} \right) ,
		\end{equation}
		where
		\begin{equation}
			T^{ph}_{i+1, i} = \left( \begin{array}{c} e^{-i \Phi_D + i \Phi_L} \\ 0 \end{array}  \begin{array}{c} 0 \\ e^{i \Phi_D - i \Phi_L} \end{array} \right) ,
		\end{equation}
$\Phi_D = \exp{(2\pi i \ n_{\text{eff}} D/\lambda)}$, $\Phi_L = \exp{(2\pi i \ n_{\text{eff}} L/\lambda)}$ and $n_{\text{eff}}$ is the effective refractive index of the resonator material.

The $N$-ring transfer matrix is the concatenated product of $T_i$ and $T^{ph}_{i+1, i}$
		\begin{equation}
		\label{NRing-InputOutput-Relation}
			T^{(N)} = \left( \begin{array}{cc} T^{(N)}_{1,1} & T^{(N)}_{1,2} \\ T^{(N)}_{2,1} & T^{(N)}_{2,2} \end{array} \right) = T_N \ T^{ph}_{N, N-1} \dots T_2 \ T^{ph}_{2, 1} \ T_1 .
		\end{equation}


	\section{Results and discussion}
	\label{SecRes&Dis}
	
In this section, we consider the case in which the incident light is pumped from the left side via port 1; thus, $S'_{+N} = 0$. The $N$-ring transfer matrix yields the amplitude of transmitted light through the system
		\begin{equation}
			S'_{-N} = \frac{ \text{det} \left( T^{(N)} \right) }{ T^{(N)}_{2, 2} } S_{+1} .
		\end{equation}

The transmittance of light through $N$ resonators is
		\begin{equation}
			T = \frac{|S'_{-N}|^2}{|S_{+1}|^2} = \frac{1}{|T^{(N)}_{2, 2}|^2} .
		\end{equation}


\subsection{The case of one resonator}

\begin{figure}[H]
			\centering
			\begin{subfigure}{0.5\textwidth}
				\includegraphics[scale=0.70]{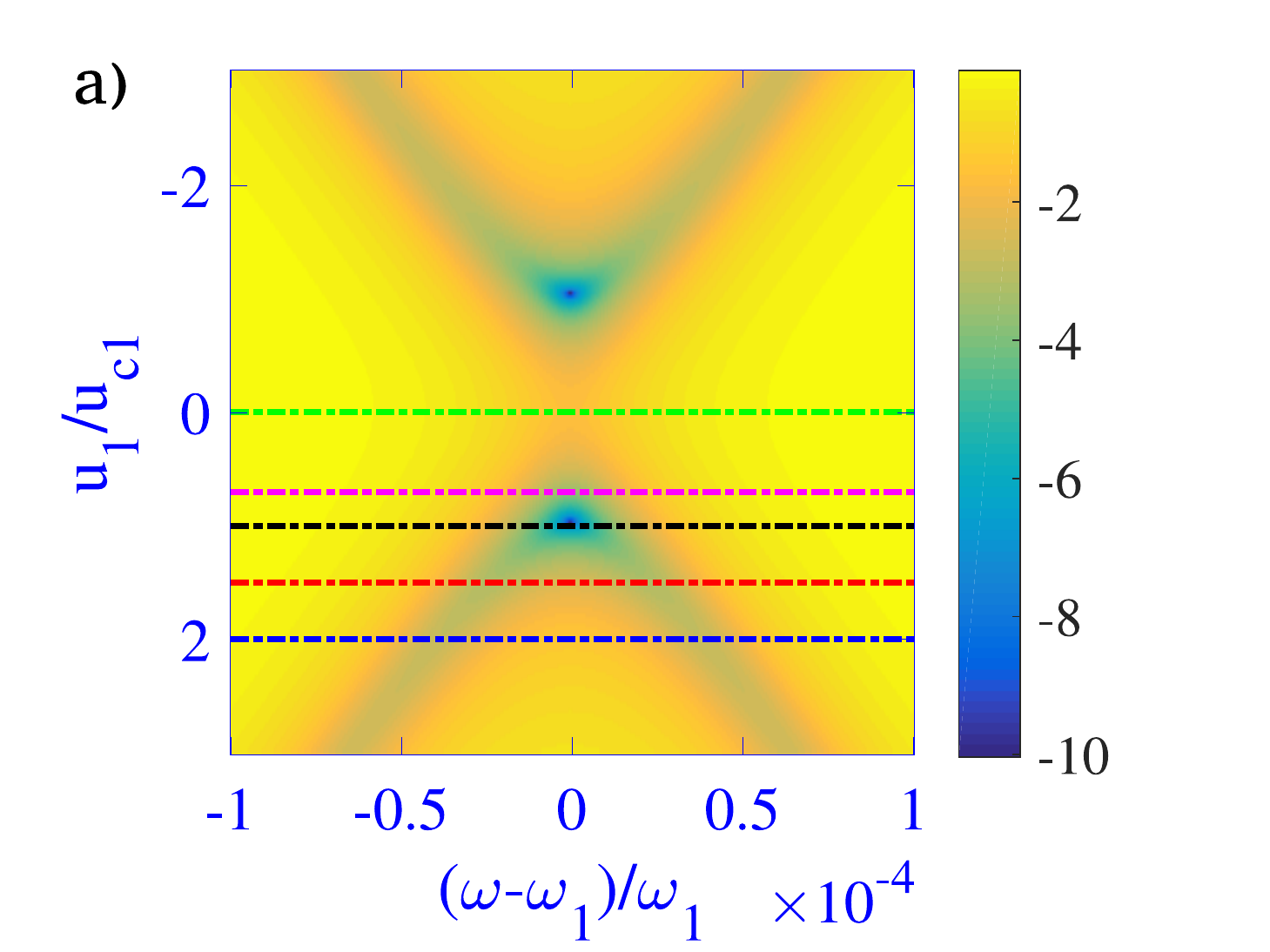}
			\end{subfigure}~ ~ ~ ~ ~ ~ ~ ~ ~ ~
			\begin{subfigure}{0.5\textwidth}
				\includegraphics[scale=0.4]{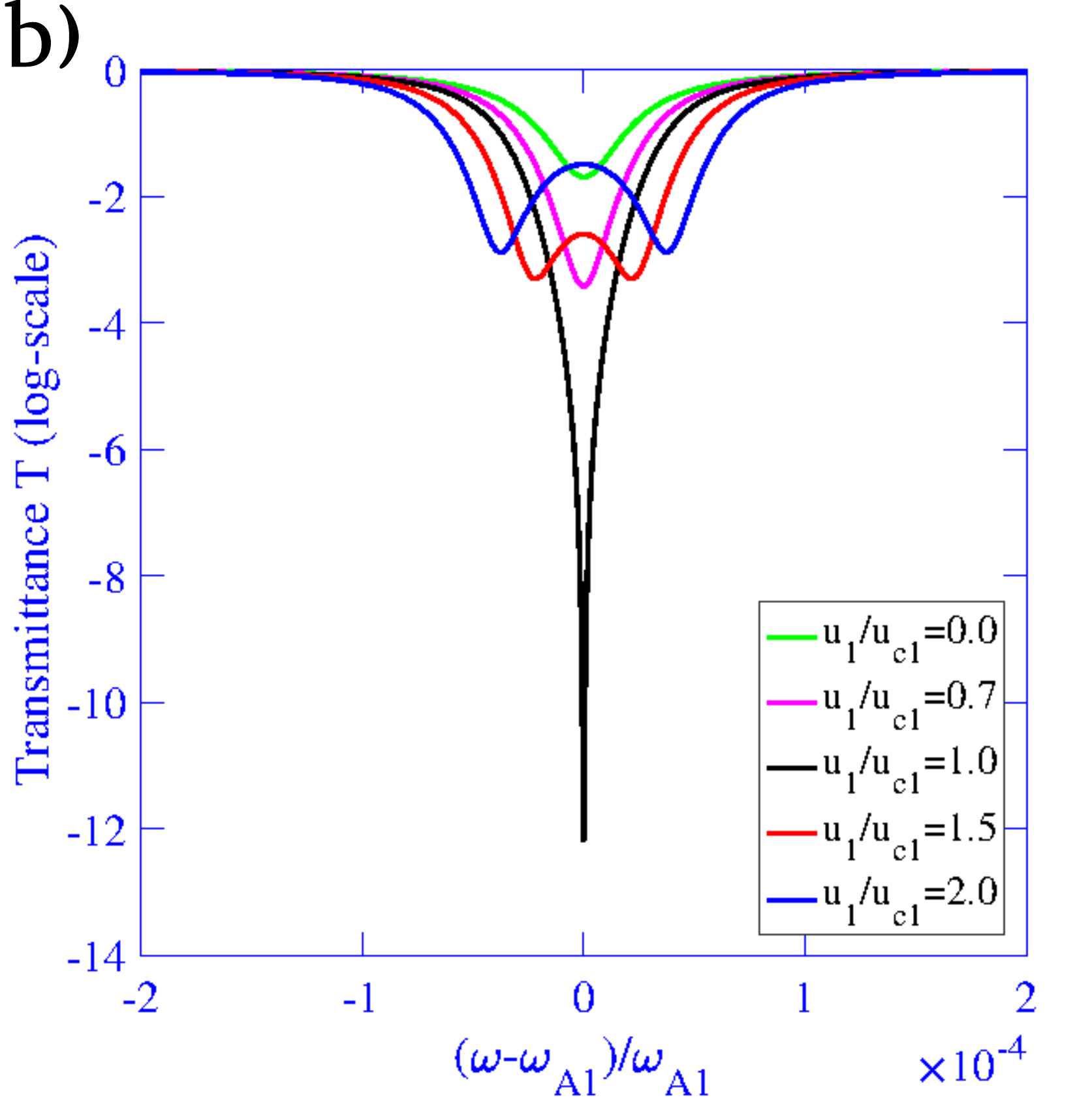}
			\end{subfigure}
			\caption{Transmission spectrum $T$ in the case of $N=1$ resonator; (a) Color map of $T(\omega, u_1)$, the colored dash lines indicate the values of $u_1$ at which the corresponding transmission spectra $T(\omega)$ are plotted on Fig. \ref{1Ring}(b); (b) $T(\omega)$ for various values of $u_1$. $\omega$: frequency of pumped light; $u_1$: mutual-mode coupling coefficient. Note the mode-splitting that occurs for $|u_1/u_{c1}| > 1$. The splitting indicates that when the incoming light is pumped at the cavity resonant frequency, the strong coupling between light and cavity modes is lost. \label{1Ring}}
		\end{figure}

		The case of one resonator is instructive. For concreteness, we let $ \lambda_{a} = \lambda_{b} = 637.7 nm $, $Q_{a,e} = Q_{b,e} = 2 \times 10^4$ and $Q_{a,i} = Q_{b,i} = 5 \times 10^4$. The choice of the central wavelength $\lambda_{a/b}$ corresponds to a zero phonon line of a nitrogen vacancy center in diamond. Ring radius $L = 1 \mu m$ of which the actual value does not affect the result for $T$. The quantity $u_{c1}$ is the critical coupling coefficient such that the splitting in the transmission spectrum occurs for $ |u| > |u_{c1}| $ (see Fig. \ref{1Ring}). This mode-splitting phenomenon has been established in \cite{Zhang:08}.

		\begin{figure}[H]
			\centering
			\begin{subfigure}{0.5\textwidth}
				\includegraphics[scale=0.70]{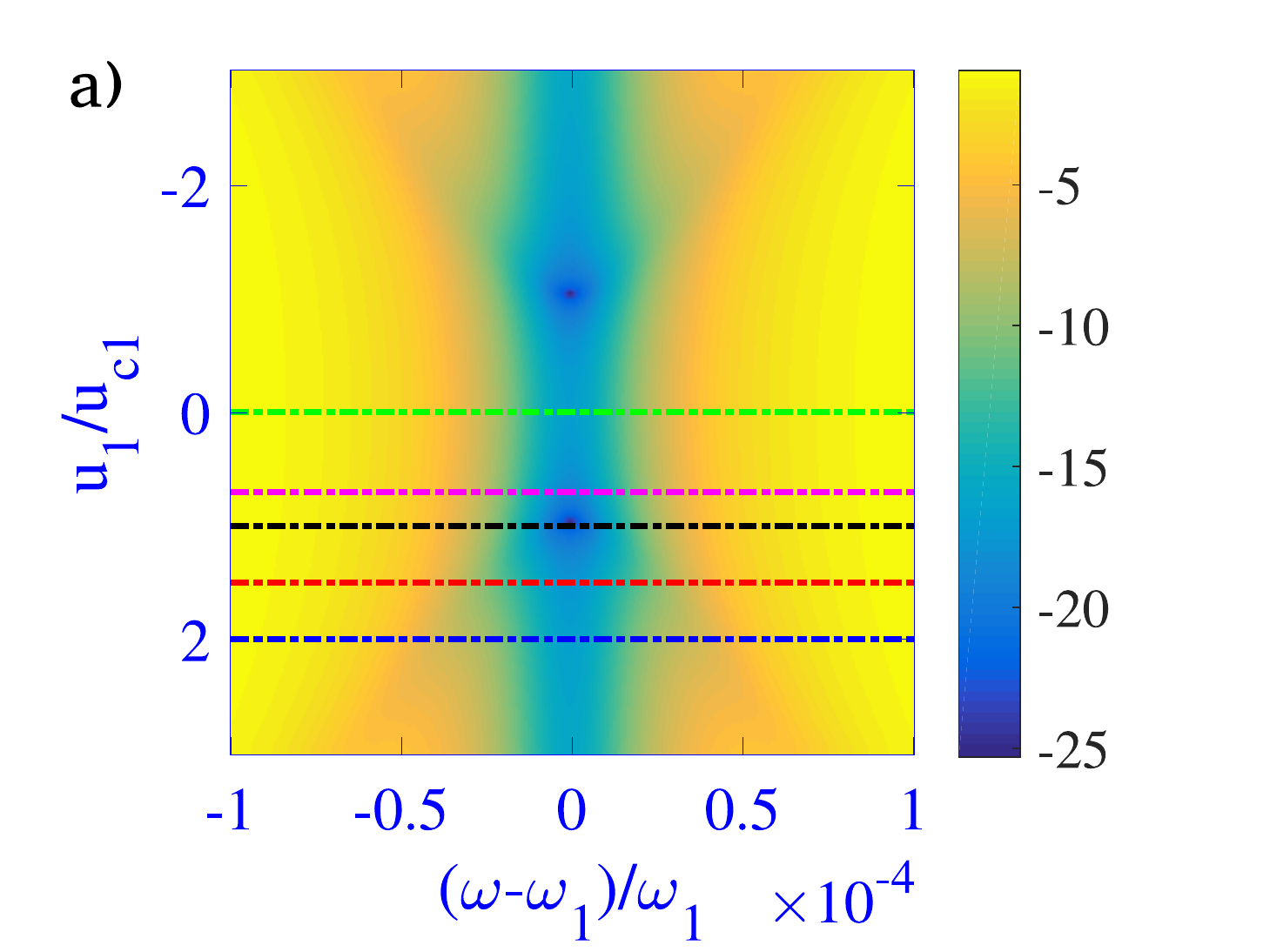}
			\end{subfigure}~ ~ ~ ~ ~
			\begin{subfigure}{0.5\textwidth}
				\includegraphics[scale=0.4]{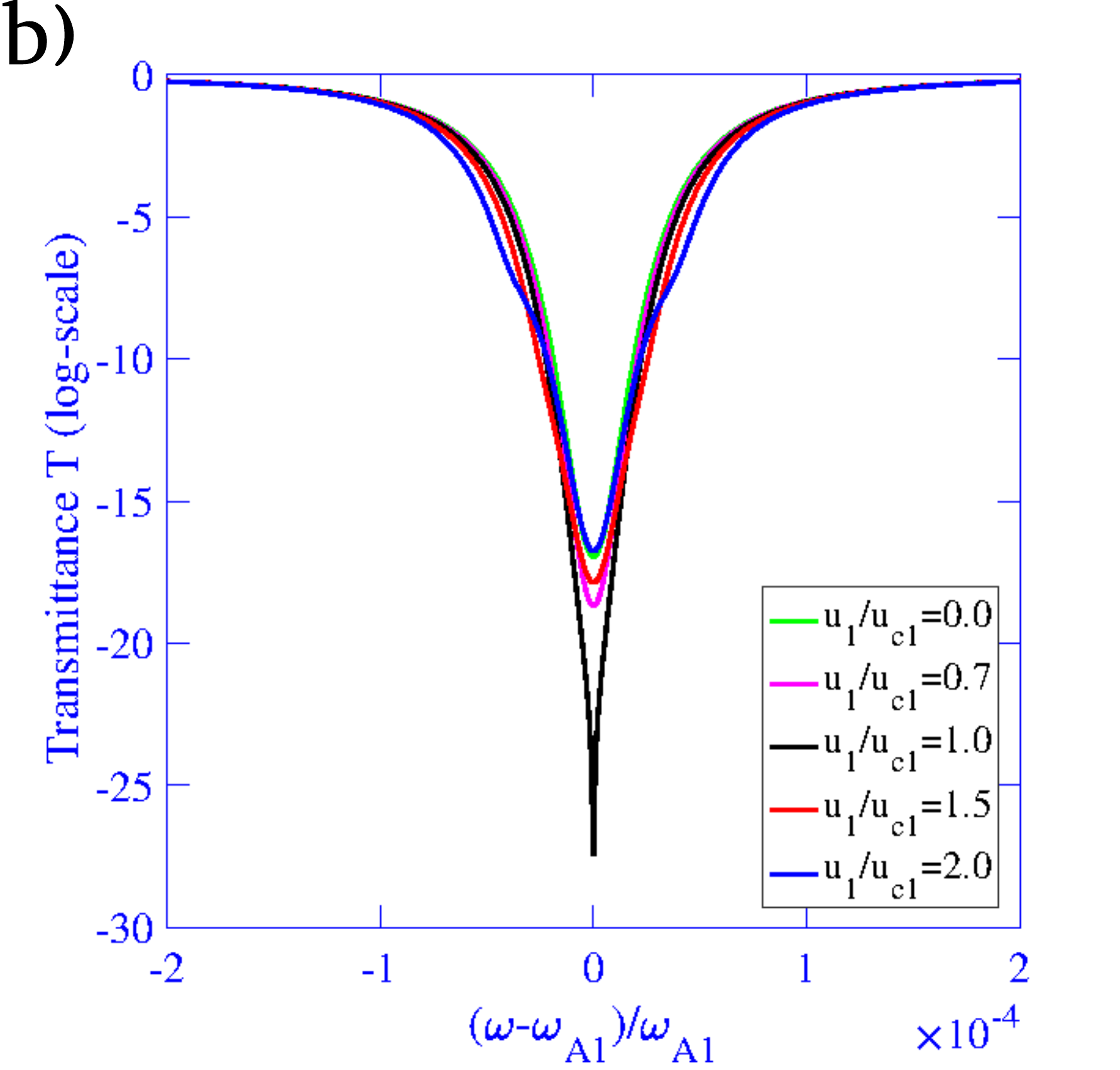}
			\end{subfigure}
			\caption{Transmission spectrum $T$ in the case of $N=10$ resonators with $u_i = 0.01 \ u_{ci}$ for $i = 2, ..., N$; (a) Color map of $T(\omega, u_1)$, the colored dash lines indicate the values of $u_1$ at which the corresponding transmission spectra $T(\omega)$ are plotted on Fig. \ref{10Ring}(b); (b) $T(\omega)$ for various values of $u_1$. $\omega$: frequency of pumped light; $u_1$: mutual-mode coupling coefficient. Even at strong coupling $|u_1/u_{c1}| > 1$, there is only one minimum at the resonant frequency $\omega = \omega_1$. The mode-splitting does not happen up to $|u_1/u_{c1}| = 3$. \label{10Ring}}
		\end{figure}
		\begin{figure}[H]
			\centering
			\begin{subfigure}{0.5\textwidth}
				\includegraphics[scale=0.70]{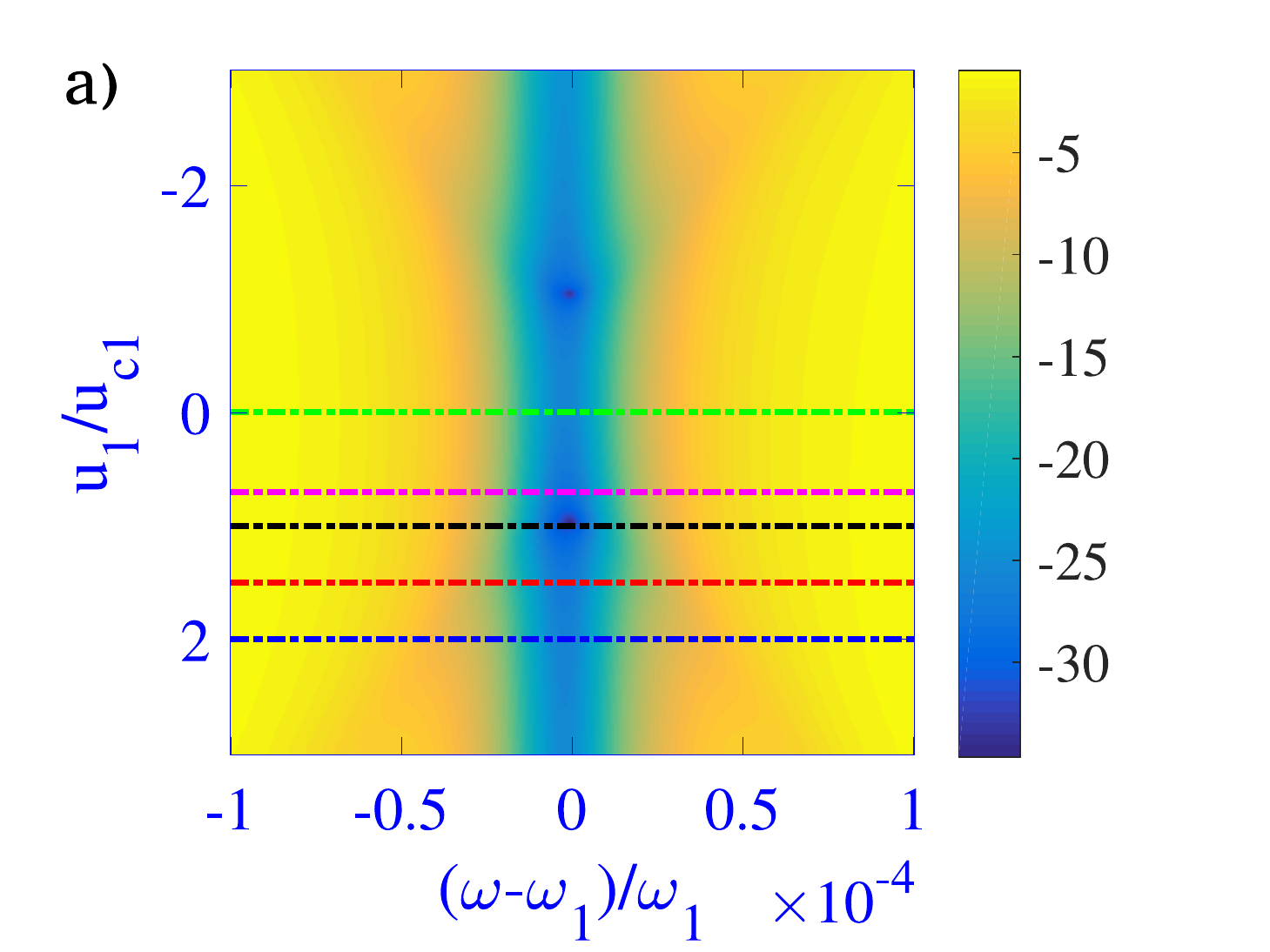}
			\end{subfigure}~ ~ ~ ~ ~
			\begin{subfigure}{0.5\textwidth}
				\includegraphics[scale=0.4]{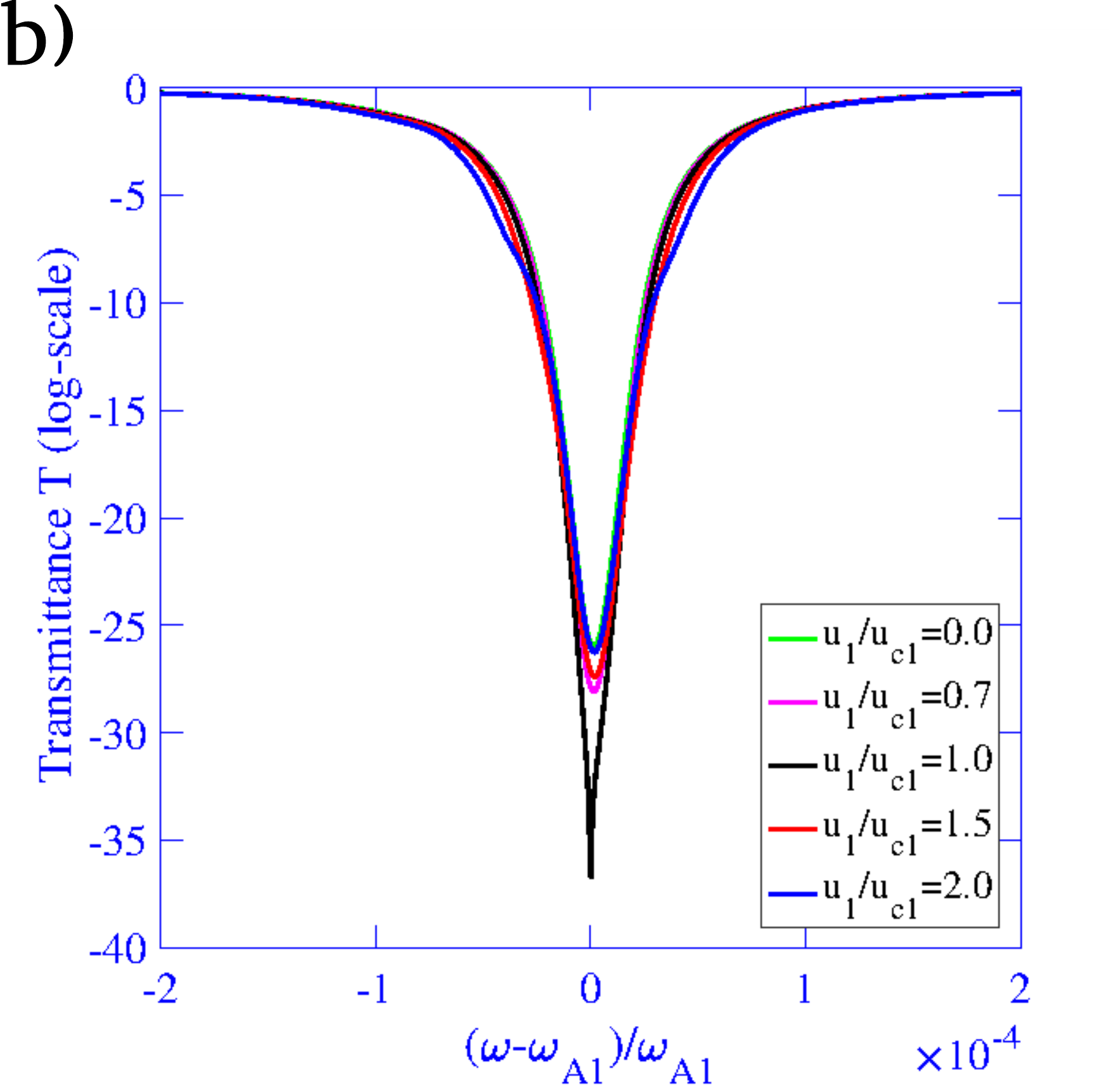}
			\end{subfigure}
			\caption{Transmission spectrum $T$ in the case of $N=10$ resonators with $u_i = 0.50 \ u_{ci}$ for $i = 2, ..., N$; (a) Color map of $T(\omega, u_1)$, the colored dash lines indicate the values of $u_1$ at which the corresponding transmission spectra $T(\omega)$ are plotted on Fig. \ref{10Ring}(b); (b) $T(\omega)$ for various values of $u_1$. $\omega$: frequency of pumped light; $u_1$: mutual-mode coupling coefficient. Again at strong coupling $|u_1/u_{c1}| > 1$, there is only one minimum at the resonant frequency $\omega = \omega_1$. The mode-splitting does not happen up to $|u_1/u_{c1}| = 3$. \label{10Ring0.5u}}
		\end{figure}

		\begin{figure}[H]
			\centering
			\begin{subfigure}{0.5\textwidth}
				\includegraphics[scale=0.70]{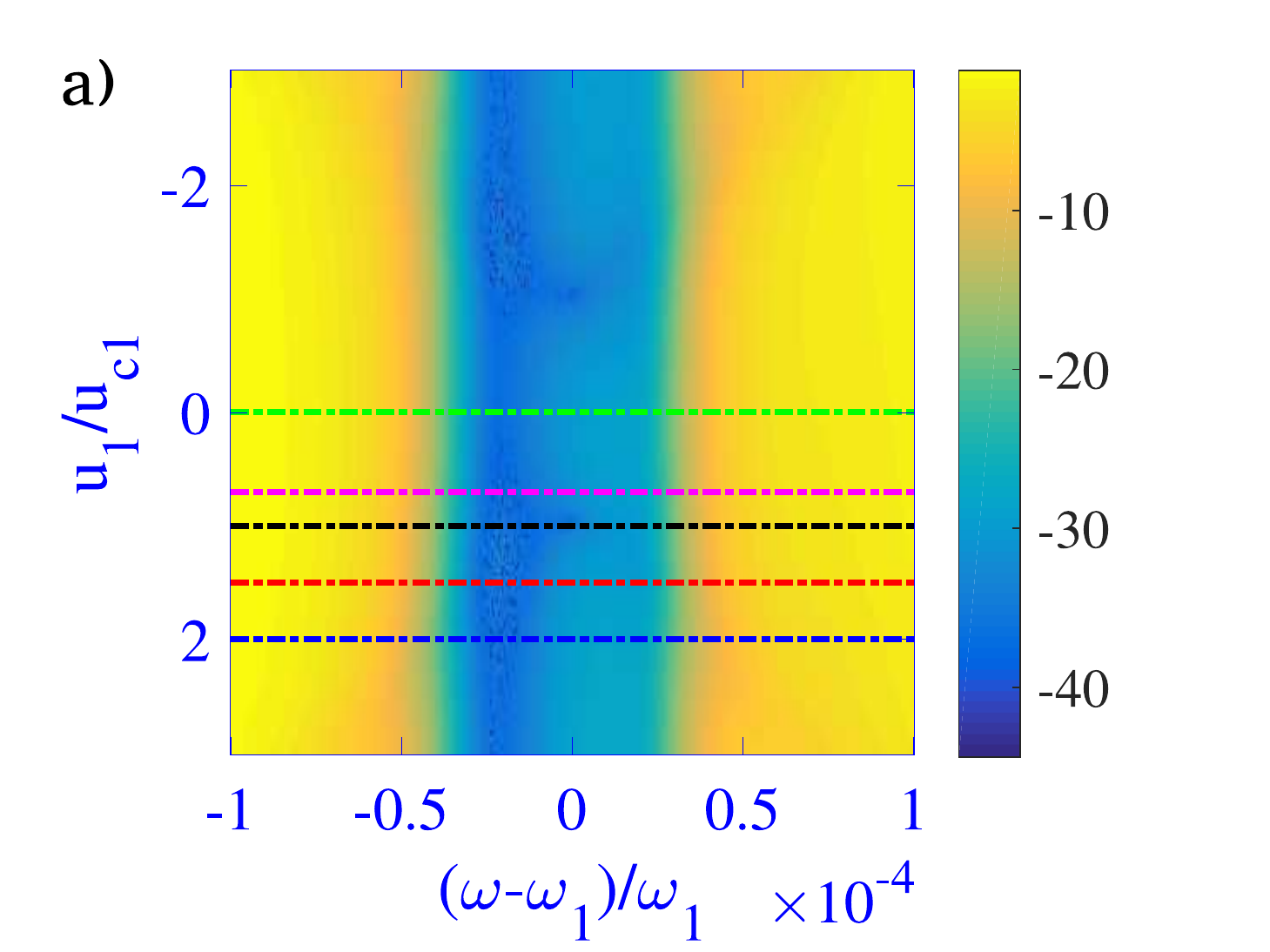}
			\end{subfigure}~ ~ ~ ~ ~
			\begin{subfigure}{0.5\textwidth}
				\includegraphics[scale=0.4]{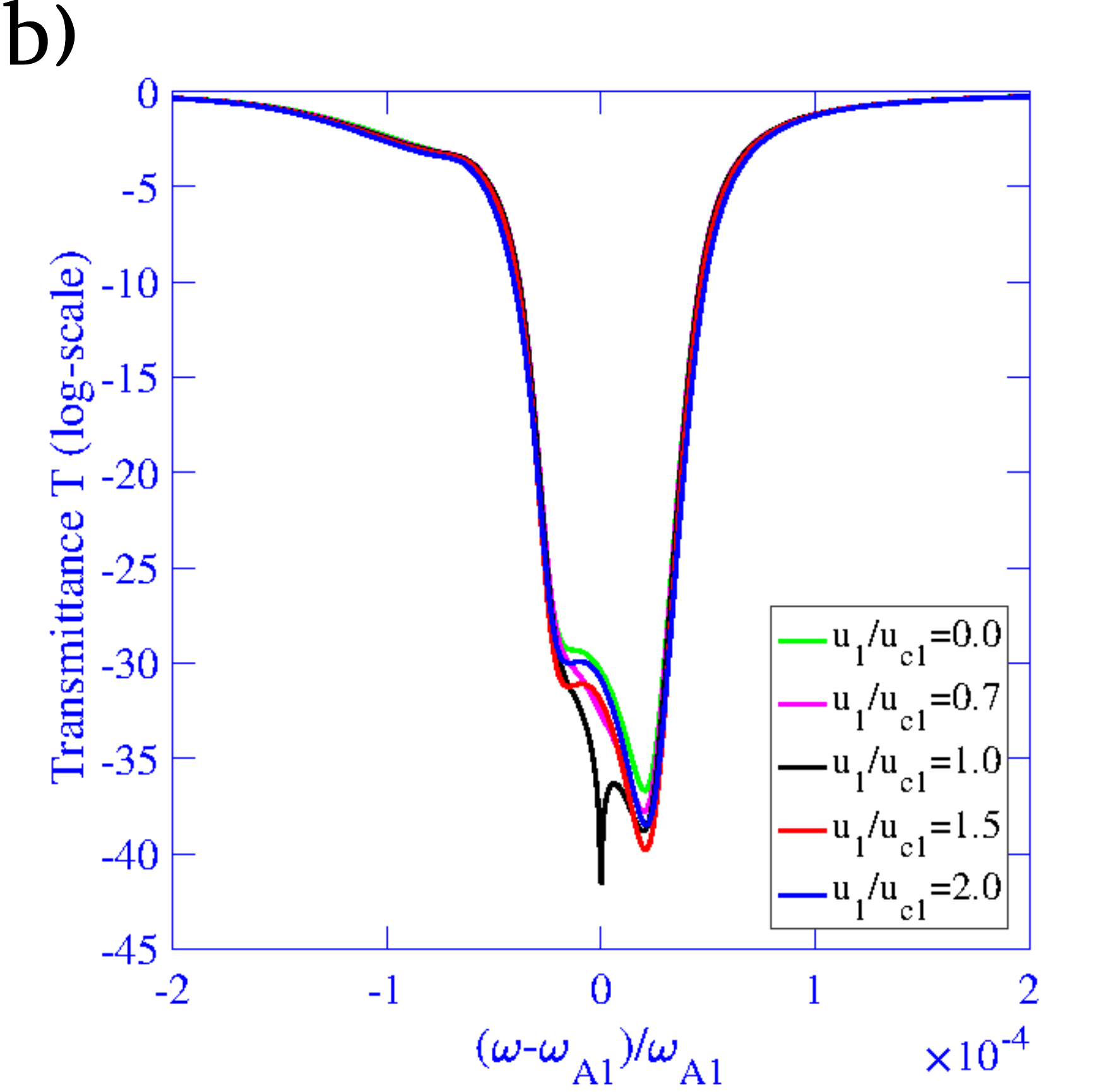}
			\end{subfigure}
			\caption{ Transmission spectrum $T$ in the case of $N=10$ resonators with $u_i = 1.50 \ u_{ci}$ for $i = 2, ..., N$; (a) Color map of $T(\omega, u_1)$, the colored dash lines indicate the values of $u_1$ at which the corresponding transmission spectra $T(\omega)$ are plotted on Fig. \ref{10Ring}(b); (b) $T(\omega)$ for various values of $u_1$. $\omega$: frequency of pumped light; $u_1$: mutual-mode coupling coefficient. Note the only minimum at the resonant frequency $\omega$, which different from $\omega_1$ in this case. The strong coupling $u_1$ does not result in additional splitting. In fact, the behaviour of $T$ when all $u_i$ are big is rather complex. We show the recovery of a single peak at $\omega$ away from $\omega_i$. \label{10Ring1.5u}}
		\end{figure}

		\begin{figure}[H]
			\centering
			\begin{subfigure}{0.5\textwidth}
				\includegraphics[scale=0.42]{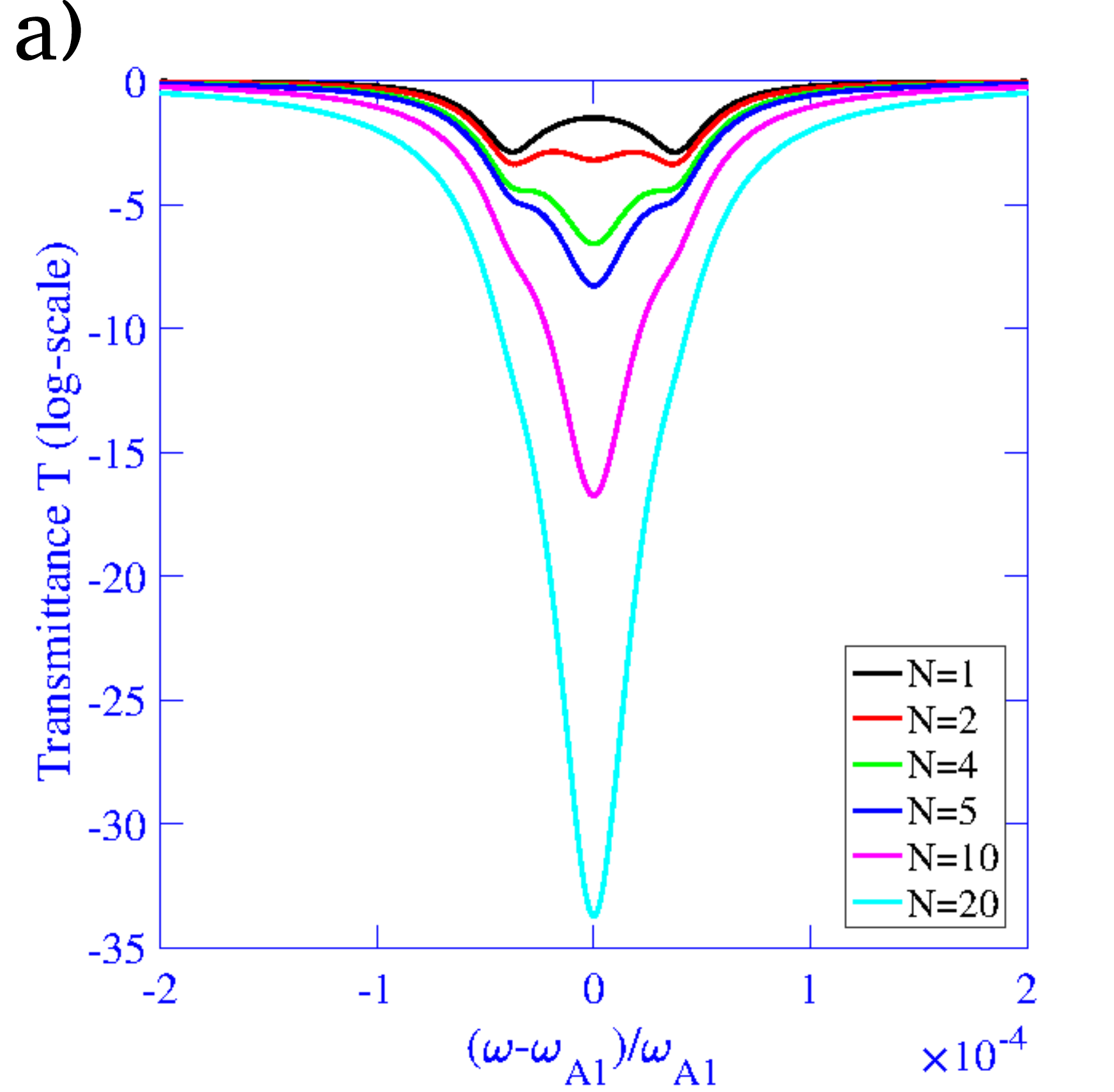}
			\end{subfigure}~ ~ ~ ~ ~
			\begin{subfigure}{0.5\textwidth}
				\includegraphics[scale=0.42]{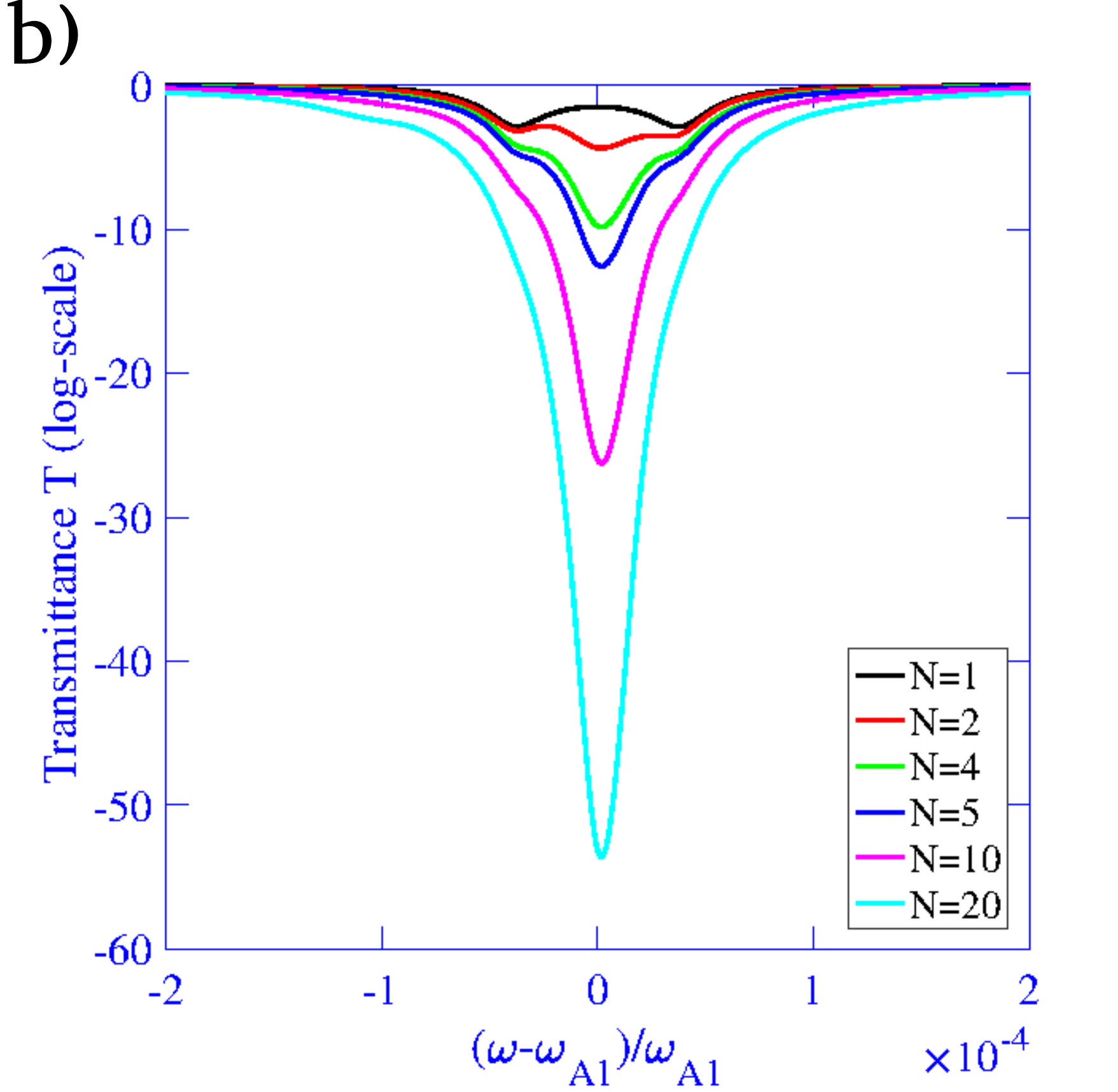}
			\end{subfigure}
			\caption{ Transmission spectrum $T$ versus $\omega$ for various values $N$ at $u_1 = 2.0 \ u_{c1}$, $\omega$: frequency of pumped light; (a) when all $u_i = 0.01 u_{ci}$ for $i = 2, \dots , N$ ; (b) when all $u_i = 0.5 u_{ci}$ for $i = 2, \dots , N$. Note the splitting gradually disappears as $N$ increases. The minimum at $\omega = \omega_1$ (when there is no splitting) in each case appears deeper and broader as $N$ becomes larger. \label{2u1_0.01_and_0.5_ui_NRing}}
		\end{figure}

		\begin{multicols}{2}
			\begin{figure}[H]
				\centering
				\includegraphics[scale=0.4]{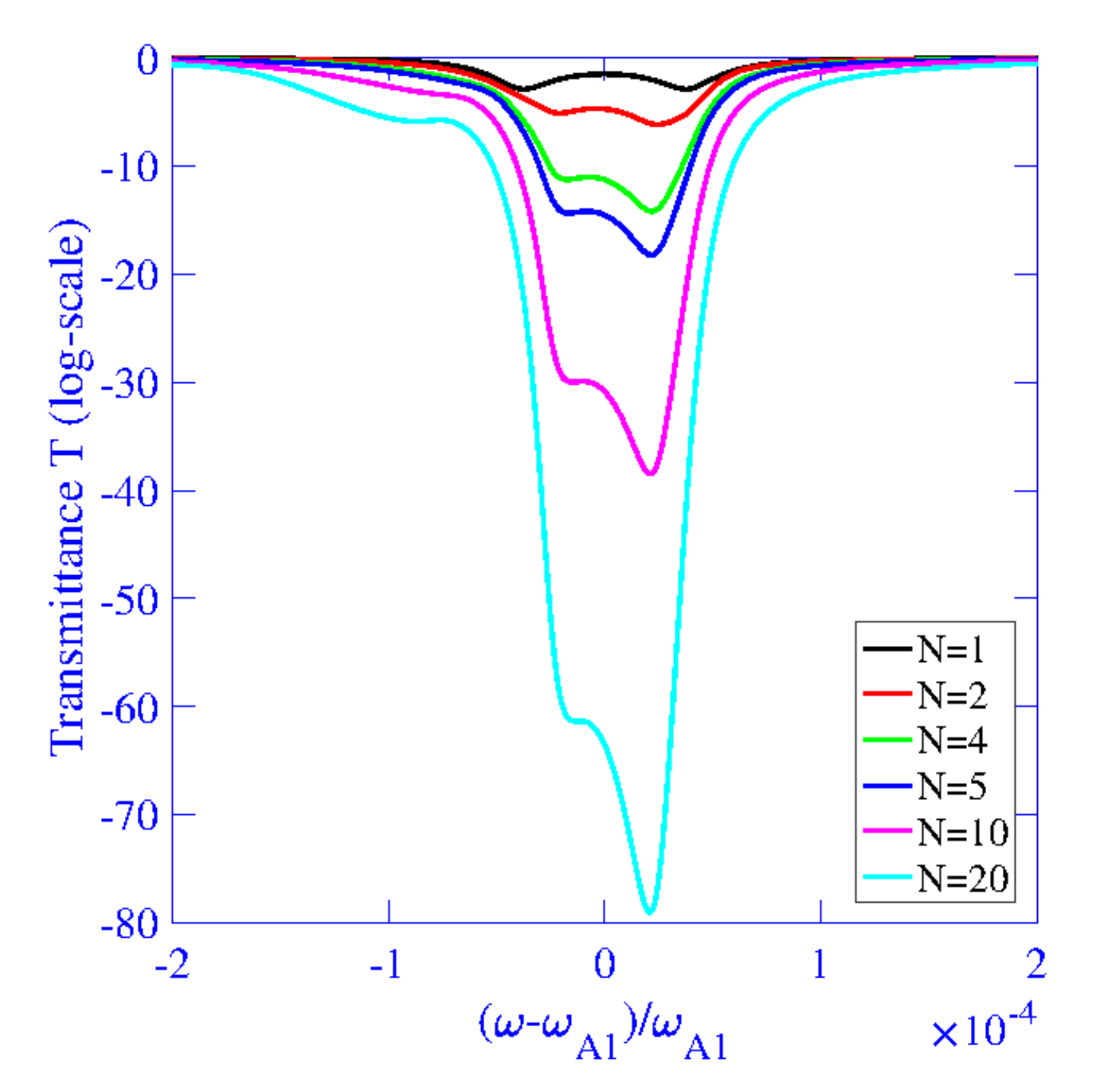}
				\caption{$T(\omega)$ for various values of $N$; $u_1 = 2.0 u_{c1}$, $u_i = 1.5 u_{ci}$ for $i=2, \dots, N$. One side dip slowly disappears and a single minimum is retained as $N$ increases. However, the other dip remains and does not happen at the original $\omega_i$.}
				\label{2u1_1.5_ui_NRing}
			\end{figure}
			\begin{figure}[H]
				\centering
				\includegraphics[scale=0.65]{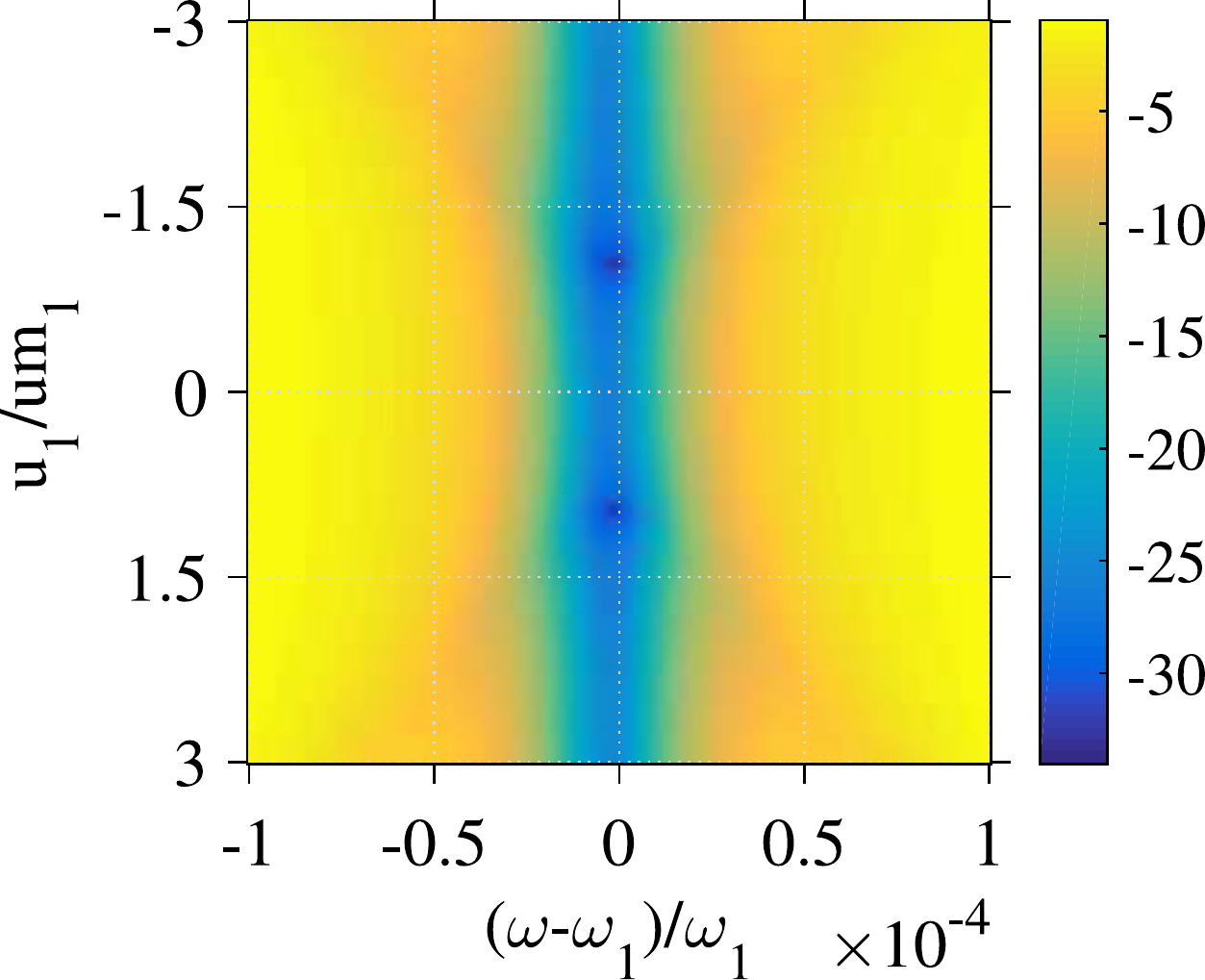}
				\caption{Color map of transmission spectrum $T(\omega, u_1)$ at $N=10$ and $u_i = 0.01 u_{ci}$, $i = 2, \dots , N$. The minimum of $T$ occurs at $\omega = \omega_1$. This feature persists even in the presence of random perturbation $\Delta D_{i+1, i}$ on $D_{i+1, i}$.}
				\label{10Ring-vary_u-Color_code-D_variation}
			\end{figure}
		\end{multicols}


\subsection{The case of $N$ resonators}

For the case of $N$ resonators,  as in the previous case, we set the $\lambda_{a_i} = \lambda_{b_i} = 637.7 nm$, $Q_{a_i,e} = Q_{b_i,e} = 2 \times 10^4$, $Q_{a_i,i} = Q_{b_i,i} = 5 \times 10^4$.
$u_i$ for $i = 2, \dots , N$ are set to three different values $u_i = 0.01 u_{ci}$, $u_i = 0.5 u_{ci}$ and $u_i = 1.5 u_{ci}$.
We take the distance between any two adjacent rings to be $D_{i+1, i} = 10 \mu m$ and ring radius to be $L = 1 \mu m$.

The transmission spectrum $T$ has only one minimum at $\omega = \omega_i$, the resonant frequency of the resonators, for $u^2_1/ u^2_{c1} \in [0, 9]$ and $u_i < u_{ci}$ for other $i$ (see Fig. \ref{10Ring} and \ref{10Ring0.5u}).
The result is robust against small perturbations $\Delta D_{i+1, i} \sim 1 \mu m$ in $D_{i+1, i}$ (Fig. \ref{10Ring-vary_u-Color_code-D_variation}).
		
\begin{figure}[h]
		\centering
		\begin{subfigure}{0.5\textwidth}
			\includegraphics[scale=0.65]{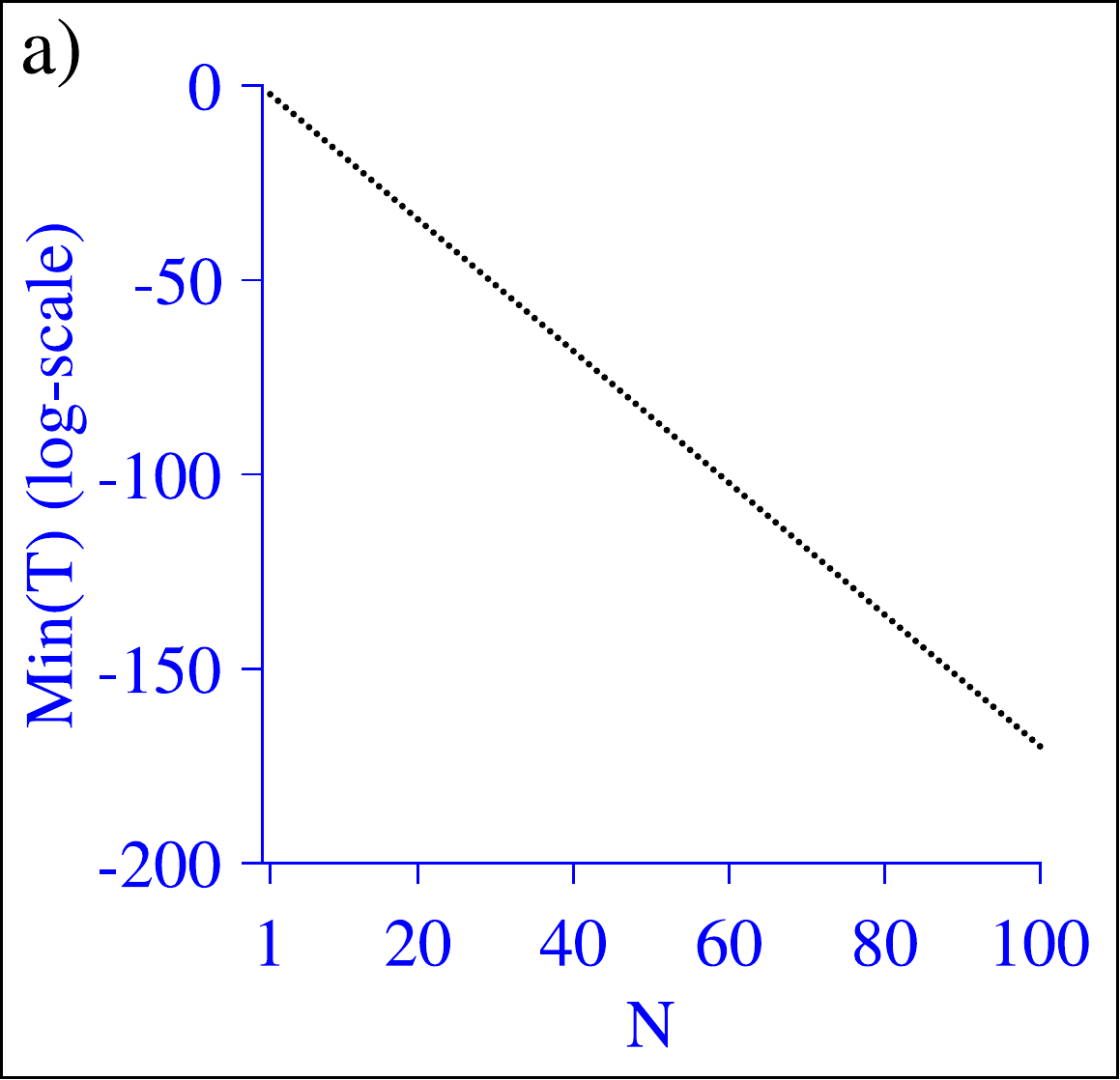}
		\end{subfigure}~ ~ ~ ~
		\begin{subfigure}{0.5\textwidth}
			\includegraphics[scale=0.65]{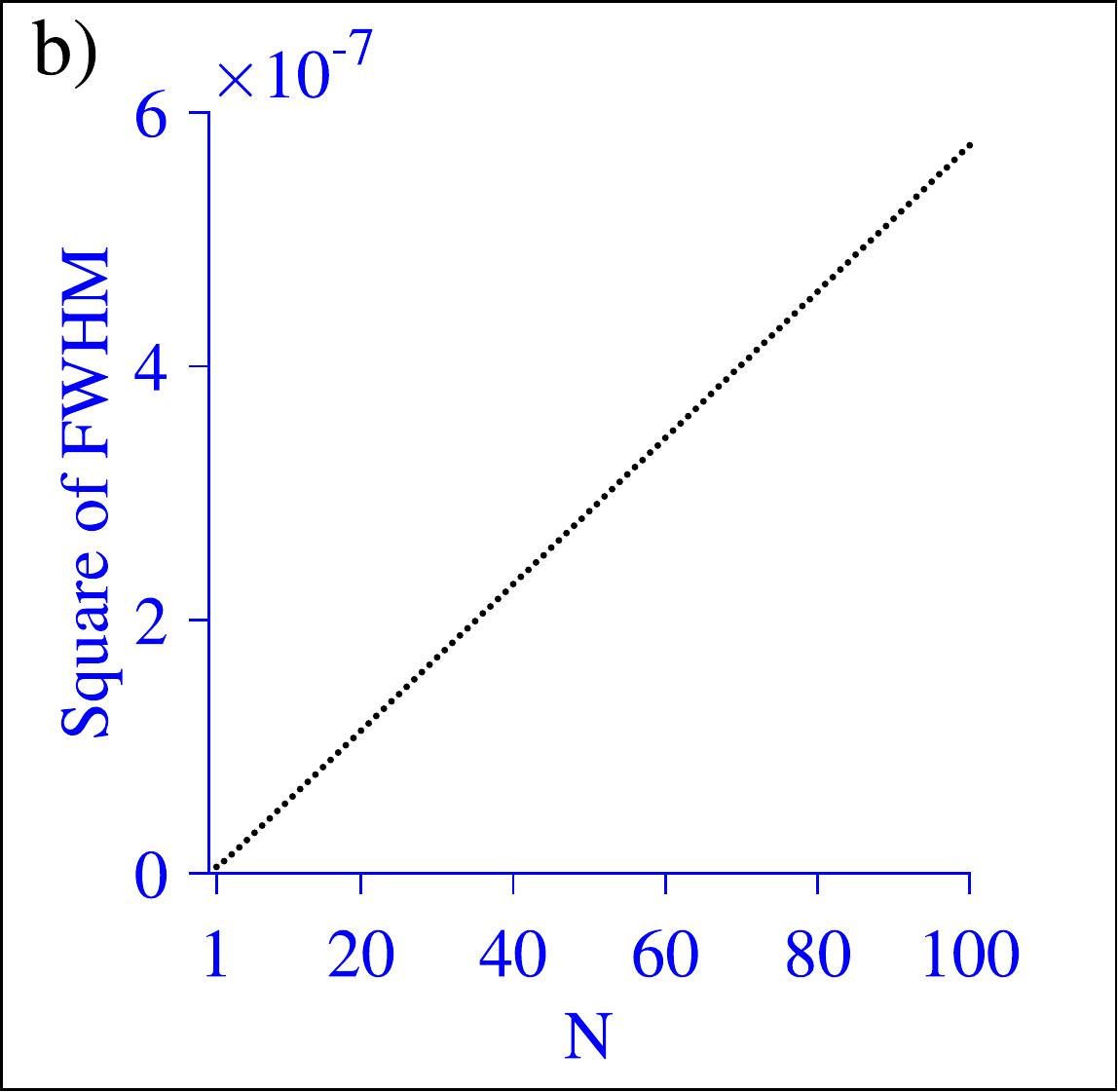}
		\end{subfigure}
		\caption{The minimum (a) and FWHM (b) of transmission spectrum $T$ versus the number of resonators $N$, $u_1 = 0.4 u_{c1}$, $u_i = 0.01 u_{ci}, \ i=2, ... , N$. The minimum decreases exponentially with $N$ while the FWHM grows as $\sqrt{N}$. The transmission spectrum becomes deeper and broader when $N$ increases. \label{TversusN}}
	\end{figure}

Intuitively, one can think of each resonator as a reflector for light. When more resonators are added, the probability of light being transmitted thus becomes lower. The coupling of light to the cavity array is enhanced as $N$ becomes larger. From  Fig. \ref{2u1_0.01_and_0.5_ui_NRing} and \ref{TversusN} (a), the system of $N$ cavities behaves like a ``giant cavity" with a broadened linewidth compared to the single-cavity case.

	One notices the following important effects of increasing $N$ on the transmission spectrum.
	\begin{enumerate}
		\item The mode-splitting that occurs at $u^2_1 > u^2_{c1}$ disappears when $N$ becomes larger enough and $u_{i} < u_{ci}$ for other rings; see Fig. \ref{10Ring}, \ref{10Ring0.5u} and \ref{2u1_0.01_and_0.5_ui_NRing}. The central dip is recovered.
		\item When the mutual mode couplind coefficients are also big in other rings, one observes that there is a broad minimum at $\omega_0$ that slightly deivates from $\omega_i$ and is independent of $u_1.$ This observation may be interpreted as just a partial recovery from the mode splitting, see Fig. \ref{10Ring1.5u} and \ref{2u1_1.5_ui_NRing}.
		\item The transmission dip at $\omega = \omega_{a_i} = \omega_{b_i}$ gets deeper, as shown in Fig. \ref{TversusN}(a) for $u_i < u_{ci}$. One observes the exponential dependence of the minimum of $T$ on $N$, which is a direct implication of Eq. (\ref{NRing-InputOutput-Relation}).
		\item As $N$ becomes bigger, the transmission dip around the resonant frequency becomes broader. The FWHM of $T$ grows as $\sqrt{N}$ as shown in Fig. \ref{TversusN}(b). Denote $|\kappa_{b_i}|^2 = \Gamma_e$, $\Gamma_{b_i} = \Gamma = \Gamma_e + \Gamma_i$ and $\Delta \omega = \omega-\omega_{b_i}$.
			From the one-ring transfer matrix (\ref{T_i}),
			\begin{equation}
				t_i = \left| 1 - \frac{|\kappa_{b_i}|^2}{i(\omega-\omega_{b_i}) + \Gamma_{b_i} + \frac{u_i^2}{i(\omega-\omega_{a_i}) + \Gamma_{a_i}}} \right|^2 .
			\end{equation}

			In the case where $u_i \ll \Gamma$, $t_i \approx \frac{\Delta \omega^2 + \Gamma_i^2}{\Delta \omega^2 + \Gamma^2}$. Heuristically, when light travels across any resonator, a fraction of it, which equals to $\frac{\Delta \omega^2 + \Gamma_i^2}{\Delta \omega^2 + \Gamma^2}$, is transmitted. After passing through $N$ rings, the amount of transmitted light is 
			\begin{equation}
				T \approx \left( 1 - \frac{\Gamma^2 - \Gamma_i^2}{\Delta \omega^2 + \Gamma^2} \right)^N \approx 1 - N \frac{\Gamma^2 - \Gamma_i^2}{\Delta \omega^2 + \Gamma^2} \approx \frac{\Delta \omega^2 - N \Gamma^2}{\Delta \omega^2 + \Gamma^2} \ \ \ (\Gamma_i \ll \Gamma)
			\end{equation}
			Therefore, at $\Delta \omega^2 = (2N+1) \Gamma $ the transmission $T \approx \frac{1}{2}$, which intuitively explains with the relationship in Fig. \ref{TversusN}(b).
	\end{enumerate}

	\section{Conclusion and outlook}
	\label{SecConclusion}

We have studied the effects of $N$ rings on the transmission spectrum in 1-D waveguide. Each ring behaves as a reflector for light and thus gives rise to a transmission dip at the resonant frequency of the ring. The mode splitting, which may occur at the mutual-mode coupling coefficient $u_1$ larger than $u_{c1}$--critical coupling coefficient, was reduced in the transmission spectrum for large $N$. The effect is robust against significant variations in the ring-to-ring separations $D_{i+1, i}$. When there is no mode-splitting, the transmission dip at the resonant frequency $\omega_o = \omega_{a_i} = \omega_{b_i}$ becomes deeper as the number of rings increases and exhibits exponential dependence on $N$ with its FWHM proportional to $\sqrt{N}$.

Future work will be focused on the studies of how (\textit{i}) the variation in the resonant frequencies $\omega_{i}$ of the rings, (\textit{ii}) the strong coupling $u_i, \ i>1$, in other rings and (\textit{iii}) the direct coupling of the modes $a/b_i$ to $a/b_{i \pm 1}$ affect the transmission spectrum.
The coupling between the system of arbitray $N$ resonators with quantum emitters such as cold atoms, nitrogen-vacancy centres in diamonds etc, can be another exciting subject of future study.

\pagebreak

\section*{Acknowledgement}
This work was supported by NRF-CRP14-2014-04, Engineering of a Scalable Photonics Platform for Quantum Enable Technologies.

T. P. Tan Nguyen would like to thank T. T. Ha Do for her help with the access to Matlab and her comments on the figures.



\bibliographystyle{unsrt}
\bibliography{Bibliography}





\end{document}